\pdfoutput=1
\documentclass[fleqn,usenatbib]{mnras}

\usepackage[T1]{fontenc}

\DeclareRobustCommand{\VAN}[3]{#2}
\let\VANthebibliography\thebibliography
\def\thebibliography{\DeclareRobustCommand{\VAN}[3]{##3}\VANthebibliography}

\usepackage{graphicx}	
\usepackage{amsmath}	
\usepackage{amssymb}	

\newcommand{\beq}{\begin{equation}}
\newcommand{\eeq}{\end{equation}}
\newcommand{\Mdot}{\dot{M}}
\newcommand{\hii}{H~{\sc II}~}

\newcommand{\Mo}{\mbox{M$_{\odot}$}}
\newcommand{\Moy}{\mbox{M$_{\odot}$ yr$^{-1}$}}

\title[Common Envelope Shaping of Planetary Nebulae. IV]{Common Envelope Shaping of Planetary Nebulae. IV. \\
From Proto-planetary  to Planetary Nebula}

\author[Garc\'{\i}a-Segura et al.]{
Guillermo Garc\'{\i}a-Segura,$^{1}$\thanks{E-mail: ggs@astrosen.unam.mx (GGS)}
Ronald E. Taam,$^{2}$
and Paul M. Ricker$^{3}$
\\
$^{1}$Instituto de Astronom\'{\i}a, Universidad Nacional Aut\'onoma
de M\'exico, Km. 107 Carr. Tijuana-Ensenada, 22860, Ensenada, B. C., Mexico\\
$^{2}$Center for Interdisciplinary Exploration and Research in Astrophysics (CIERA), 
Department of Physics and Astronomy, 
\\ Northwestern University, 
2145 Sheridan Road, Evanston, IL 60208, USA\\
$^{3}$Department of Astronomy, University of Illinois, 1002 W. Green St., Urbana, IL 61801, USA }

\date{Accepted XXX. Received YYY; in original form ZZZ}

\pubyear{2022}

\begin{document}
\label{firstpage}
\pagerange{\pageref{firstpage}--\pageref{lastpage}}
\maketitle

\begin{abstract}

We present 2D hydrodynamical simulations of the transition of a proto-planetary nebula 
to  a
planetary nebula  for central stars in binary systems that have undergone a  common envelope event. 
After 1,000 yr of magnetically driven dynamics (proto-planetary nebula phase), a 
 line-driven stellar wind is introduced into the  computational domain and the expansion
of  the
nebula is simulated for another 10,000 yr,  including  the effects of stellar photoionization.
 In this study we consider central stars with main sequence (final) masses of 1 (0.569) and 2.5 (0.677) \Mo, together with a 0.6 \Mo main sequence companion.
Extremely bipolar, narrow-waisted proto-planetary nebulae result 
in bipolar planetary nebulae,
while the rest of the shapes 
mainly evolve  into elliptical planetary nebulae.
The initial magnetic  field's  effects on the collimated structures, such as jets, tend to
disappear in most of the cases,  leaving behind the remnants of those features in only
a few cases. 
Equatorial zones fragmented mainly by photoionization ( 1 \Mo progenitors), result in ``necklace'' 
structures made of  cometary clumps  aligned with the radiation field.
On the other hand, fragmentation by photoionization and shocked wind 
( 2.5 \Mo progenitors) 
give rise to the formation of multiple clumps in the latitudinal direction, which remain 
within the lobes, close to the center, 
which are immersed and surrounded by hot shocked gas,  not necessarily aligned with 
the radiation field. These results 
reveal that the fragmentation process has a dependence on the stellar mass progenitor.
This fragmentation is made possible by the distribution of gas in the previous post-common envelope proto-planetary nebula 
as sculpted by the action of the jets.


\end{abstract}

\begin{keywords}
Stars: Evolution --Stars: Rotation --Stars: AGB and 
post-AGB --Stars: binaries  --ISM: planetary nebulae --ISM: individual
(Hubble 5, NGC 6302, NGC 2440)

\end{keywords}

\section{Introduction}

Planetary Nebulae (PNe) are the final, colorful results  of the evolution of low and 
intermediate mass  stars, allowing these stars to become white dwarfs (Kwitter 2022 and references therein ). 
Classical stellar evolution theory combined with hydrodynamical studies
(Mellema 1993)  explained the formation of 
PNe in  first approximation, but they could not explain the number of high-density structures  in PNe 
(Gon\c calves 2001, Mizalski et al. 2009),
the existence of collimated structures in proto-planetary nebulae (PPNe; Sahai \& Trauger 1998), or the fact that 
PPNe are characterized by a greater kinetic energy than 
can be provided by radiation pressure (Bujarrabal et al 2001). In other words, classical
stellar evolution theory could not explain the majority of PPNe.
However, the binary scenario in general ( Soker 1997, De Marco 2009, Jones \& Boffin 2017) as well as common envelope evolution (CEE; Paczy\'nski 1971, Taam et al. 1978, De Marco 2009, Ivanova et al. 2013,
Garc\'{\i}a-Segura et al. 2018 (paper I), Frank et al. 2018, Garc\'{\i}a-Segura et al. 2020 (paper II), Zou et al. 2020, Garc\'{\i}a-Segura et al. 2021 (paper III), Ondratschek et al. 2022)  has been able to 
provide a  consistent  
explanation for most of  the above issues in recent years (De Marco 2009, Blackman 2022). 

Binaries and CEE  have been invoked to form bipolar  PNe in many previous studies,  including the references above, 
but they have not been  necessarily involved in the  explanation of the formation of elliptical nebulae.
Theoretically, models for the formation of elliptical nebulae used ad hoc equations in 
the past  for the asymptotic giant branch (AGB) wind distribution (Icke et al 1989, Mellema et al. 1991, Frank \& Mellema 1994) or equations based on the 
 AGB stellar rotation (Garc\'{\i}a-Segura et al. 1999). 
%
%
In this article we show that elliptical nebulae are also a natural consequence of CEE. 

Our previous articles (Paper II and Paper III), based on 
the CEE computations by Ricker \& Taam (2012),
explored the idea that the gas that  remains gravitationally bound after a CEE 
event  is able to form a magnetized, accretion  circumbinary disk  that  launches  a magnetized 
wind and jet that  can form PPNe  with large kinetic energies, consistent  
with the kinematic analysis of the observations of molecular CO gas (Bujarrabal et al. 2001).
In this article, we  continue those calculations by incorporating a classic, 
line-driven, stellar wind as well as 
 the effect of photoionization, to compute the further evolution of the PN phase.

This article is a continuation of Paper III and is structured as follows.  The
numerical scheme and physical assumptions are described in \S~2. 
The results of the numerical simulations are presented in \S~3.   
Finally, we discuss the numerical results in \S~4 and provide the main conclusions in the last section.

\section{Numerical methods and physical assumptions}


\begin{figure*}
    \centering
    \includegraphics[width=0.9\linewidth]{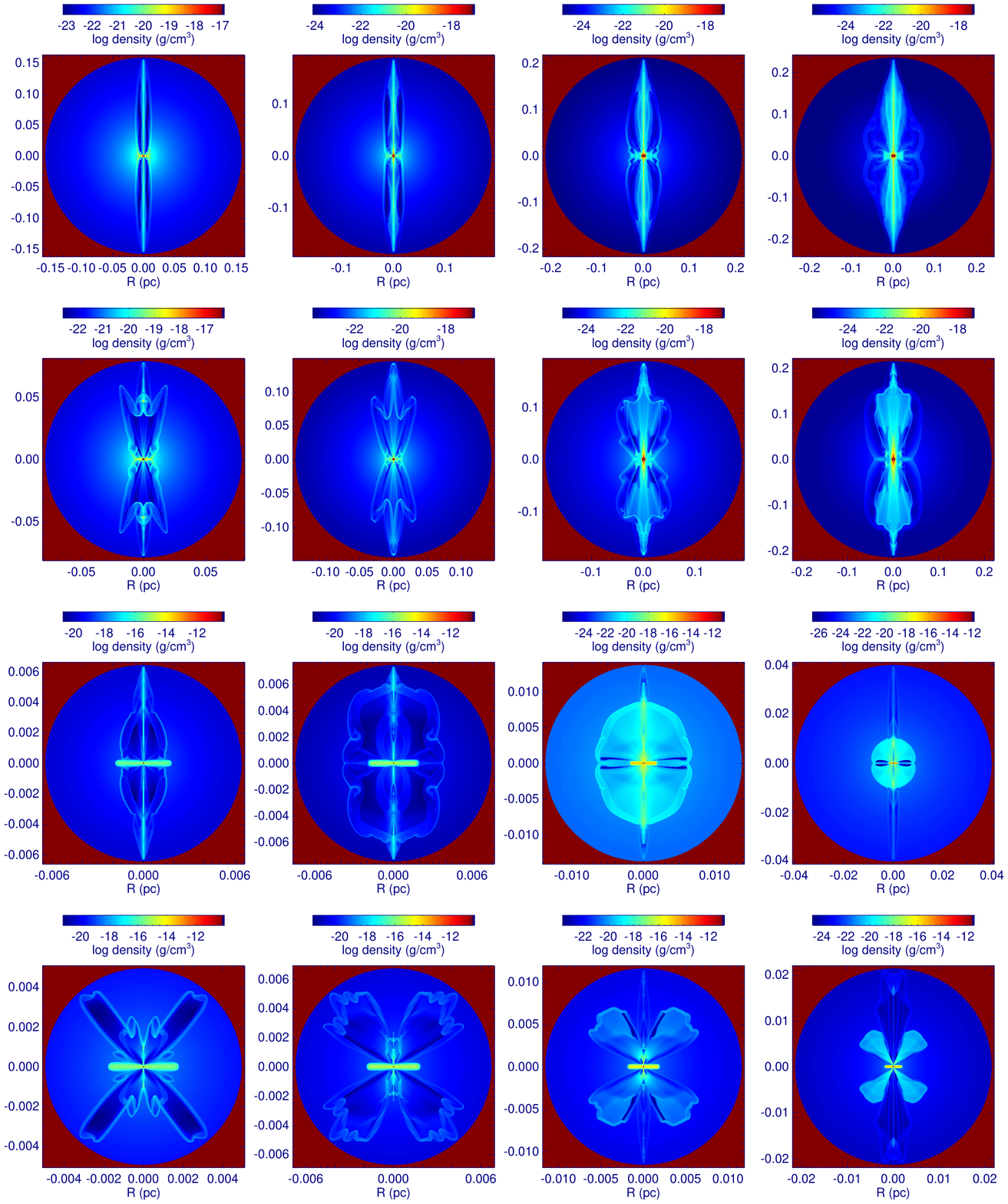}
    \caption{Logarithm of density for  PPN models B, C, E, F (top to bottom)  as shown in Paper III, with logarithm of AGB mass-loss rates  in \Moy of 
      -6, -7, -8, -9  (from left to right).  The order is the same as in Table 1.}
    \label{F1}
\end{figure*}

The numerical simulations have been performed using the magneto-hydrodynamic code
ZEUS-3D (Version 3.4), developed by M. L. Norman and the Laboratory for
Computational Astrophysics. It is a finite-difference, fully explicit,
Eulerian code descended from the code described in Stone \& Norman (1992) and
in Clarke (1996). A simple approximation is used to derive the location of the ionization
front in this study (Garc\'{\i}a-Segura \& Franco 1996),
assuming that ionization equilibrium applies at all times. 
The models include the Raymond \& Smith (1977) cooling curve above $10^4$ K.
For temperatures below $10^4$ K, the cooling curve given by  MacDonald \& Bailey (1981) is
applied. 

 We assume axisymmetry. The  2-dimensional computational grid is in spherical coordinates and consists of 
$800 \times 200$ equidistant zones in $r$ and 
$\theta$ respectively, with an angular extent of $90^{\circ}$, and
initial radial extent  in line with the models of  Paper III (see below).
A self-expanding grid technique  has been employed in order to allow the spatial coordinates to grow by several orders of 
magnitude.

\begin{table}
\caption{Initial models from Paper III}
\begin{tabular}{lcc}
\hline
Model & Log $\Mdot_{\rm AGB}$ ($\Moy$) & $B_{\phi}$\\
\hline
Fast wind launching \\
B6, B7, B8, B9 & -6, -7, -8, -9  & $1/r$  \\
C6, C7, C8, C9 & -6, -7, -8, -9  & $1/r^{2}$ \\
\hline
Slow wind launching  \\
E6, E7, E8, E9  & -6, -7, -8, -9  &  $1/r$     \\
F6, F7, F8, F9  & -6, -7, -8, -9  & $1/r^{2}$  \\
\hline
\end{tabular}
\end{table}

The  injection of mass and momentum into the computational grid representing the stellar wind is treated as an inner boundary condition,   covering the two innermost
radial zones, and is taken from Villaver et al. (2002), based on the stellar
evolution models  for zero-age main sequence (ZAMS) masses  1 \Mo and 2.5 \Mo from Vassiliadis \& Wood (1994).

For the outer, expanding  boundary, we use mass-loss rates between 
$10^{-6}$ to $10^{-9} $ \Moy similar to Paper III.
External to the  AGB wind, the interstellar medium density is set to 0.01 ${\rm cm}^{-3}$, 
which corresponds to densities external to the Galactic spiral arms (Villaver et al. 2012).  The AGB wind velocity is assumed to be 10~km~s$^{-1}$. 

As input for the starting conditions we adopt the computed  PPN  models from Paper III, displayed in 
Figure 1 and listed in Table 1. The snapshots  in Figure 1 are taken at 1,000 years from the start of the CEE event
for the case of models B and C, and 350 years for models E and F.
 As a brief summary, Papers II and III explored the further evolution of the ejected envelope computed in a CEE event by Ricker \& Taam (2012), the formation of a circumbinary disk by the gravitationally bound gas, the magnetic launching of winds or jets by the circumbinary disk and the formation of PPNe by those computed jets. The computed PPNe were allowed to evolve for a total of 1,000 years, and these are the  ones that we use here 
  as initial conditions.  In this article we now assume that the central star has reached a $T_{\rm eff}$=10,000 K,
and  photoionization starts to play a role. This properly defines the beginning of a PN, and we set our clock to time=0 in the rest of the article. At the same $T_{\rm eff}$=10,000 K, the central star now has a line-driven wind, which 
we inject in the previously computed PPNe.
Models B and C  were obtained by a fast, magnetized wind  launched 
from a circumbinary disk in which the binary had an orbital period of 2 days,
while models E and F correspond to a slow, magnetized wind   that was launched  from a binary with an 
orbital period of 5 days.

The difference between models B and C, and  between models E and F, was the treatment of the toroidal magnetic
field at the expanding, inner boundary. Models B and E assumed that the field decays as 
$1/r$, while models C and F assumed a $1/r^{2}$ behaviour. 

During the remainder of this article, we will label the models according to 
Table 1  for their input model  and 
 the mass of the central star; for example, models B6-2.5 and  B7-2.5, or models B6-1.0 and B7-1.0.

\section{Results}

%

\begin{figure*}
    \centering
    \includegraphics[width=0.9\linewidth]{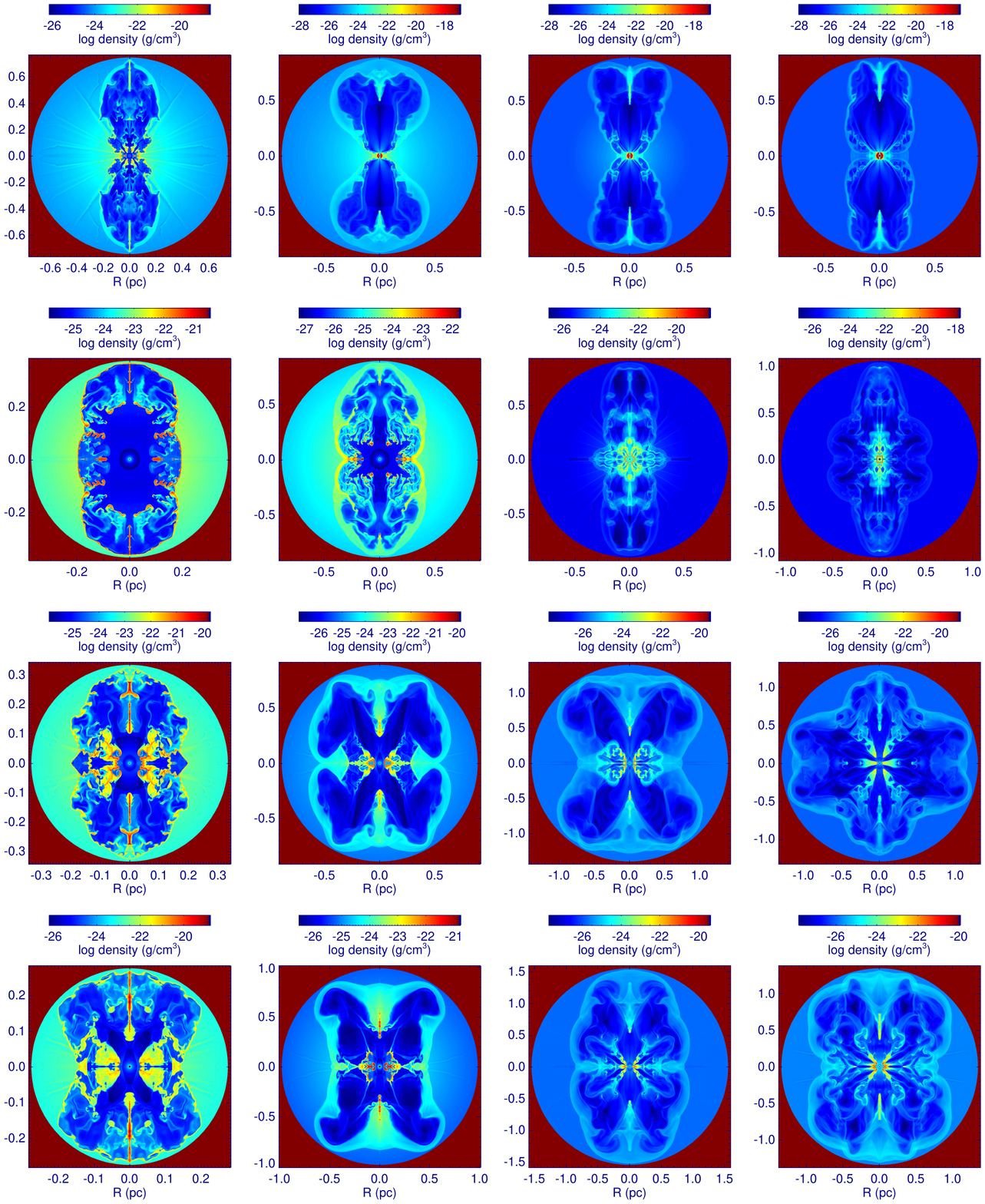}
    \caption{Logarithm of density for  PN models 
     B, C, E, F (top to bottom), with logarithm of AGB mass-loss rates  in \Moy of 
      -6, -7, -8, -9  (from left to right)  and a 2.5 \Mo  \ (ZAMS) central star, after
    3,000 yr of evolution. The order is the same as in Table 1.}
    \label{F2}
\end{figure*}


\begin{figure*}
    \centering
    \includegraphics[width=0.9\linewidth]{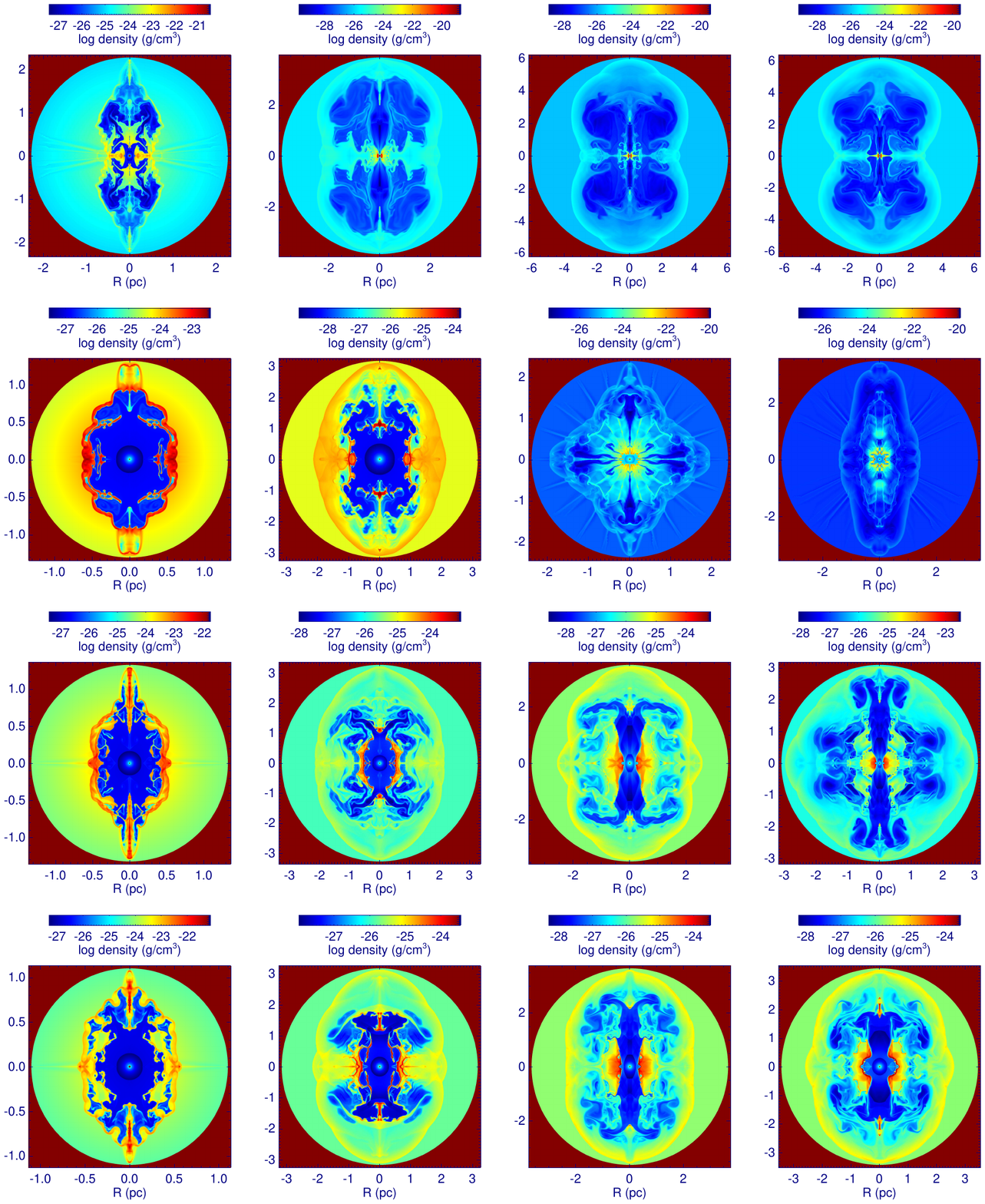}
    \caption{Logarithm of density for  PN models 
     B, C, E, F (top to bottom), with logarithm of AGB mass-loss rates in \Moy of 
      -6, -7, -8, -9  (from left to right)
     and a 2.5 \Mo  \ (ZAMS) central star, after
    10,000 yr of evolution. The order is the same as in Table 1.}
    \label{F3}
\end{figure*}

For this study we have run a total of 32 numerical simulations for a period of 10,000 years,  which is a large fraction of the typical age  of the observed PNe  (Gonz\'alez-Santamar\'{\i}a et al. 2021). 

The first 16 computations assume a central, evolving star of
0.677-\Mo , which  corresponds to a post-AGB star with a initial mass at ZAMS of 2.5 \Mo.
This set of numerical simulations  represents an average of PNe progenitors with intermediate mass  (between 2 and 3.5 \Mo). High-mass progenitors would be above 3.5 \Mo. 

The  other 16 models  assume a central post-AGB star of 0.569 \Mo, which  descends  from a star of 1.0 \Mo \, at ZAMS. This 
last set  represents the low-mass progenitor population of PNe  (below 2 \Mo).

In all cases, the companion is a  0.6 \Mo \,  main sequence star with neither a
stellar wind nor ionizing photons. 

The initial conditions for each simulation are taken from Paper III and are used for both  sets of models,
assuming that the PPN phase, in which the magnetized wind is launched from the circumbinary disk,
is similar for all of them, independently of the progenitor mass. 

 In the next subsections, the results for both stellar models will be displayed in several figures where the density scale bar is different in each  panel, with each plot scaled to its  minimum and  maximum values. Although this could be rather confusing for the reader, it is necessary because the maximum gives valuable information regarding the densities of the clumps that are observed
in each model, especially because those clumps usually are neutral and are
a guide to the densities expected for future ${\rm H}_2$  observations, as the James Webb Space Telescope has just  shown (De Marco et al. 2022, in preparation).

\subsection{Results for  the 2.5 \Mo \, stellar  model }

The first 16 computations are shown in Figure 2 at 3,000 yr and Figure 3 at 10,000 yr. 
To understand the evolution of the computed shapes it is necessary to review 
first the evolution of the central star.  

 The figures 3 and 4 in Villaver et al. (2002)
show the evolution of the stellar wind kinetic energy  and the ionizing photons
respectively. For the case of 2.5 \Mo, the wind becomes evident 
at 600 yr  after the star reaches ${\rm T_{eff} = 10,000}$ K, reaching a maximum in its kinetic energy at 2,500 yr. Subsequently,  the wind's momentum and kinetic energy rapidly decline. On the other hand, the number of ionizing photons reaches its maximum at 1,000 yr
with a fast decline at 3,000 yr. This is because the blueward evolution of the central 
star in the HR diagram is fast. As a consequence of the rapid evolution,
the wind and the ionizing photons drive the dynamics of the nebula simultaneously.

The snapshots in Figure 2 (3,000 yr) show the computed nebulae at a time near the the peak of
their wind kinetic energy, so the action of the winds  has been prominent  in forming large
lobes. For example, models B6-2.5 to B9-2.5 on the upper row of Figure 2, 
which started as  ``narrow jet'' type PPNe (extremely bipolar) in Figure 1, evolve  into narrow waist bipolar PNe, and  later into bipolar PNe with large lobes in Figure 3 (10,000 yr).  
Models C6-2.5 and C7-2.5 have a short bipolar phase 
(Figure 2) and result in
elliptical PNe (Figure 3).

 The shape of the outer or forward shock dictates the description of the nebula.  In particular, most 
of the nebulae end up as elliptical nebulae in Figure 3, except for  models B (first row in Figures 1 to 3). 
However, the bright and dense  central parts, as  in the case of
model F8-2.5, can be  confused with a bipolar shape.

All models are optically thick to the ionizing radiation at the early phases, 
and  the transition to  the optically thin regime occurs as soon as the radiation increases
(Figure 4 in Villaver et al. 2002).  On the other hand,
the transition from an optically thin to optically thick regime occurs when the ionizing radiation declines during the evolution  along the white dwarf cooling track.
The appearance of spiky features  in the density snapshots  is a consequence of the optically 
thick clumps in which the ionization front is trapped locally. The shadows
behind the clumps are cold, 
neutral, and of
lower thermal pressure, producing 
an accumulation of gas from the surrounding photoionized region. 

It is also evident that 
fragmentation can occur either by photoionization or by the hot shocked gas. 
The process giving rise to the fragmentation will be described in more detail
in the discussion section.

\subsection{Results for  the 1.0 \Mo \, stellar  model }

\begin{figure*}

        \includegraphics[width=0.9\linewidth]{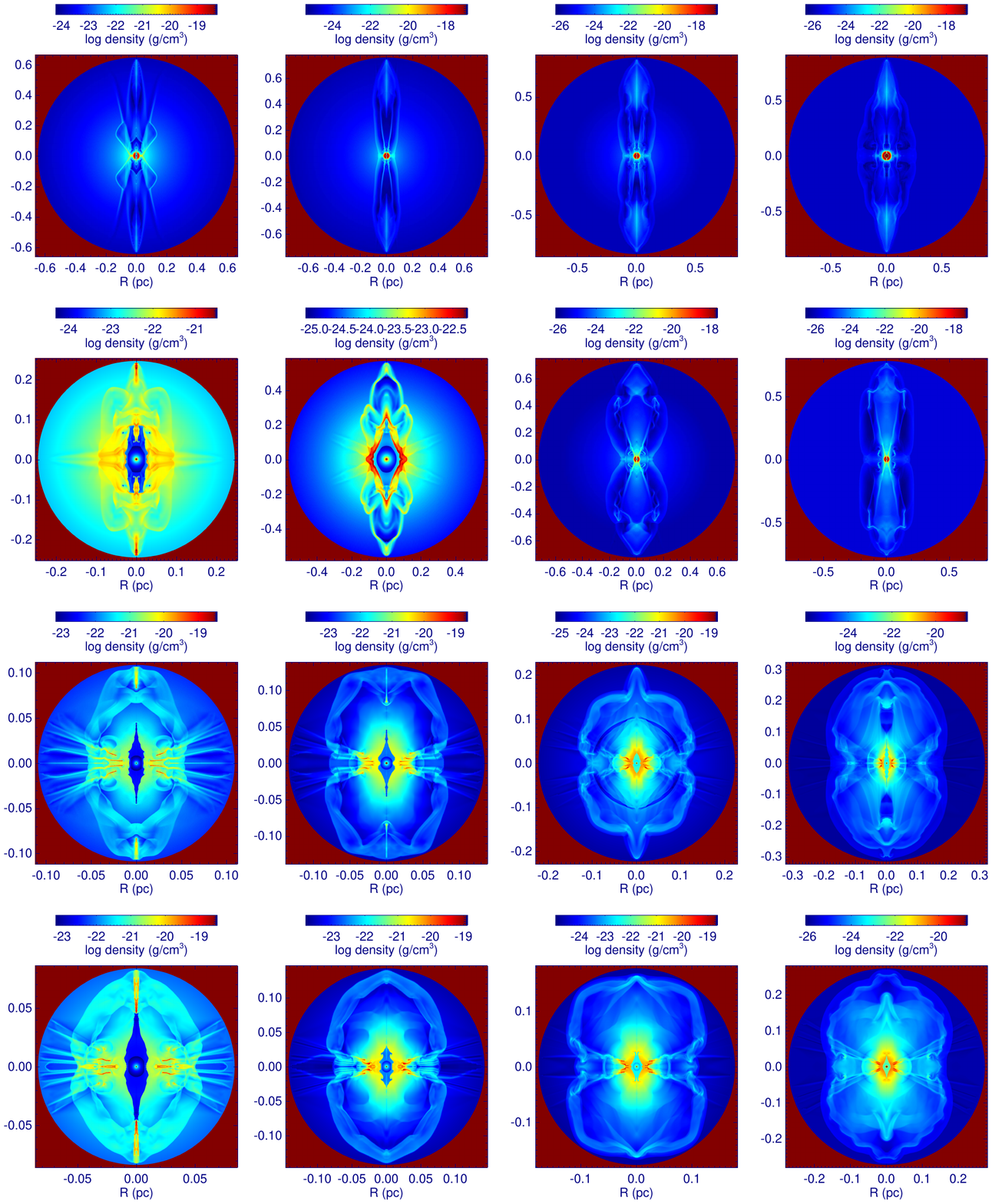}
    \caption{Logarithm of density for  PN models 
     B, C, E, F (top to bottom), with logarithm of AGB mass-loss rates  in \Moy of 
      -6, -7, -8, -9 (from left to right)  and a 1.0 \Mo  \ (ZAMS) central star, at
    3,000 yr of the evolution. The order is the same as in Table 1.}
    \label{F4}
\end{figure*}

\begin{figure*}

        \includegraphics[width=0.9\linewidth]{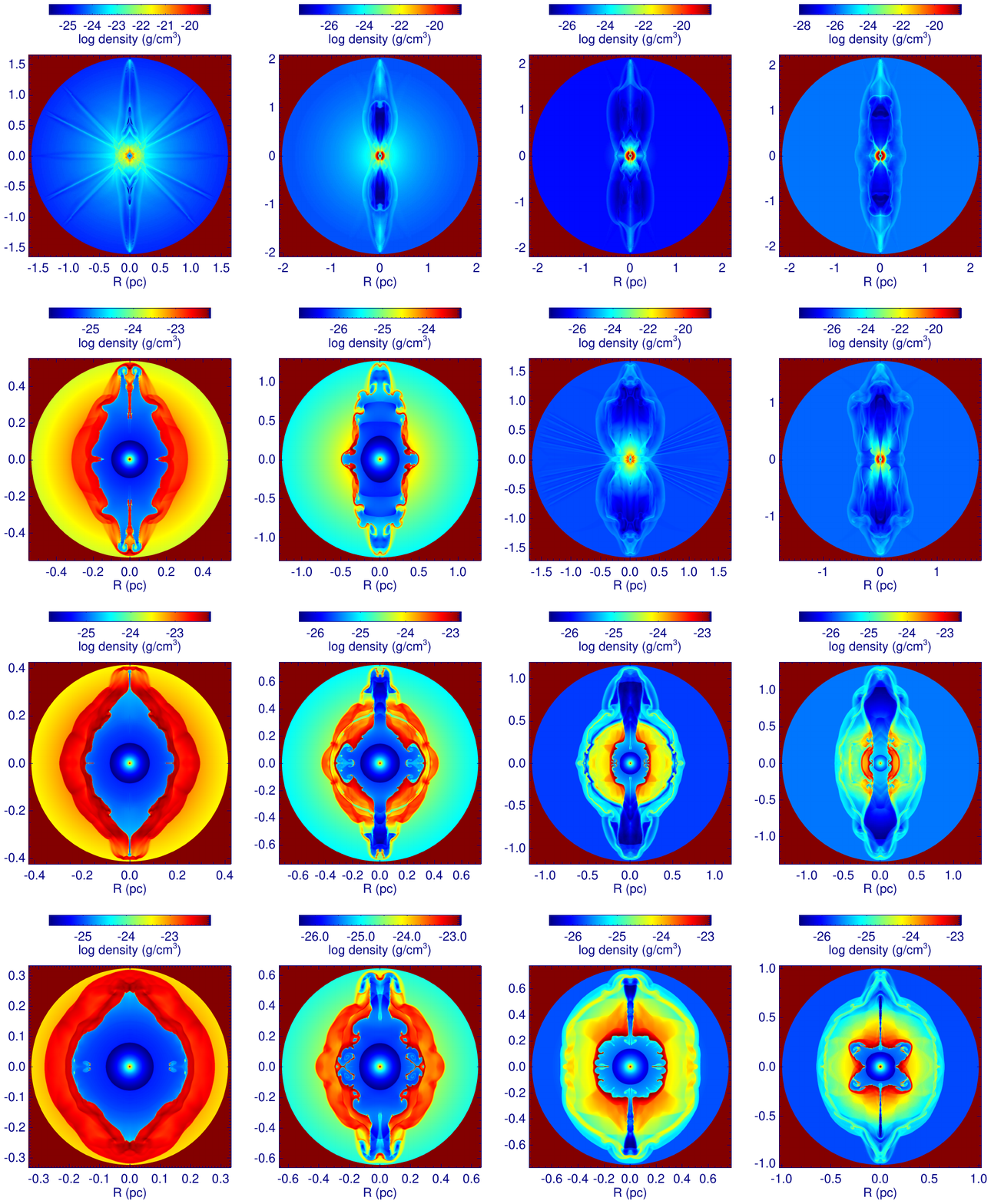}
    \caption{Logarithm of density for  PN models 
     B, C, E, F (top to bottom), with logarithm of AGB mass-loss rates  in \Moy of 
      -6, -7, -8, -9 (from left to right)  and a 1.0 \Mo  \ (ZAMS) central star, at
    10,000 yr of the evolution. The order is the same as in Table 1.}
    \label{F5}
\end{figure*}

The second set of 16 computations  is shown in  Figures 4 (3,000 yr) and 5 (10,000 yr).
It is also important to review the history of the wind from the central star for this case.
The history of the wind kinetic energy plotted in figure 3 of Villaver et al. (2002) 
shows that the wind starts to become dynamically dominant at 10,000 yr (just at the end of our computations),
reaching a maximum at 28,000 yr. On the other hand, the number of ionizing photons increases during
the total time of our computation (figure 4 of Villaver et al. 2002), and reaches a maximum
at 20,000 yr (not seen in  their  figure 4). 
According to this scenario, the  nebular dynamics are initially driven  mainly by the expansion of the \hii  region,
while the action of the wind 
becomes important later  on. For these reasons, the hot shocked gas regions have
very small volumes at 3,000 yr (Figure 4), while they begin
to be dominant at 10,000 yr (Figure 5).

 As in the computations  for the 2.5\Mo\, star, we found that the PPNe evolve into bipolar and elliptical PNe.
However, the bipolar nebulae have a different structure
since they do not form large lobes, but instead
are closer
in shapes to the original PPNe, with a very narrow waist and very elongated morphology. Their polar expansion 
velocity is also smaller, by up to a factor of 3.
The elliptical shapes in Figure 5 are also 
more similar to the ones observed in different morphological catalogues 
(Schwarz, Corradi \& Melnick 1992; Manchado et al. 1996).

Figures 4 and 5 reveal that some remnants of the PPN jets still survive, specifically for the case of models
B6-1.0 to B9-1.0. 

Particularly interesting is the equatorial fragmentation of models E-1.0 and F-1.0 in Figure 4, producing 
a large number of cometary clumps.  These might appear as ``necklace'' structures if one could resolve  them
in the $\phi$ direction in 3 dimensions.  Although they in principle could form equatorial rings, the
ionization front is very efficient in fragmenting the gas into individual clumps, since density fluctuations occurs in all spatial directions (Garc\'{\i}a-Segura et al. 2006). 

The cometary clumps (dark red in Figure 4) are neutral and
have a bright photoionized head in front.
 This fragmentation is only produced by the
photoionization since the stellar winds are still very weak at 3,000 yr. 
This is a very interesting result, 
since the possibility exists
to distinguish between models with and without powerful winds, 
 allowing us to infer the mass of the central star.

\section{Discussion}

\begin{figure*}

        \includegraphics[width=0.9\linewidth]{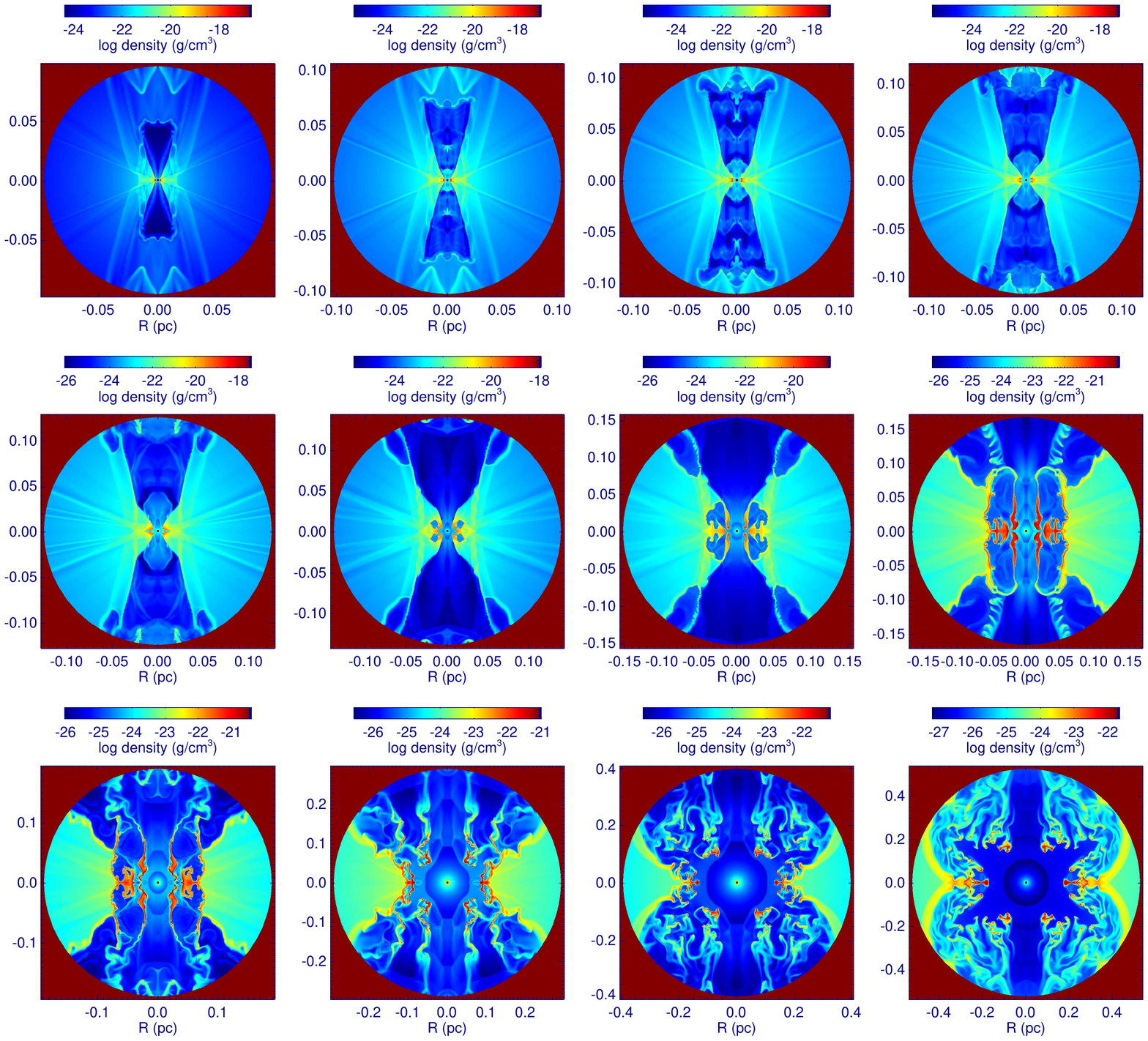}
    \caption{Logarithm of density at different snapshots (300, 400, 500, 600, 700, 900, 1,100, 1,300, 1,500, 2,000, 2,500 and 3,000 yr) of the central parts of Model C7-2.5,
    showing the fragmentation of the equatorial region. The last frame is also displayed in Figure 2.}
    \label{F6}
\end{figure*}

\begin{figure}

        \includegraphics[width=\columnwidth]{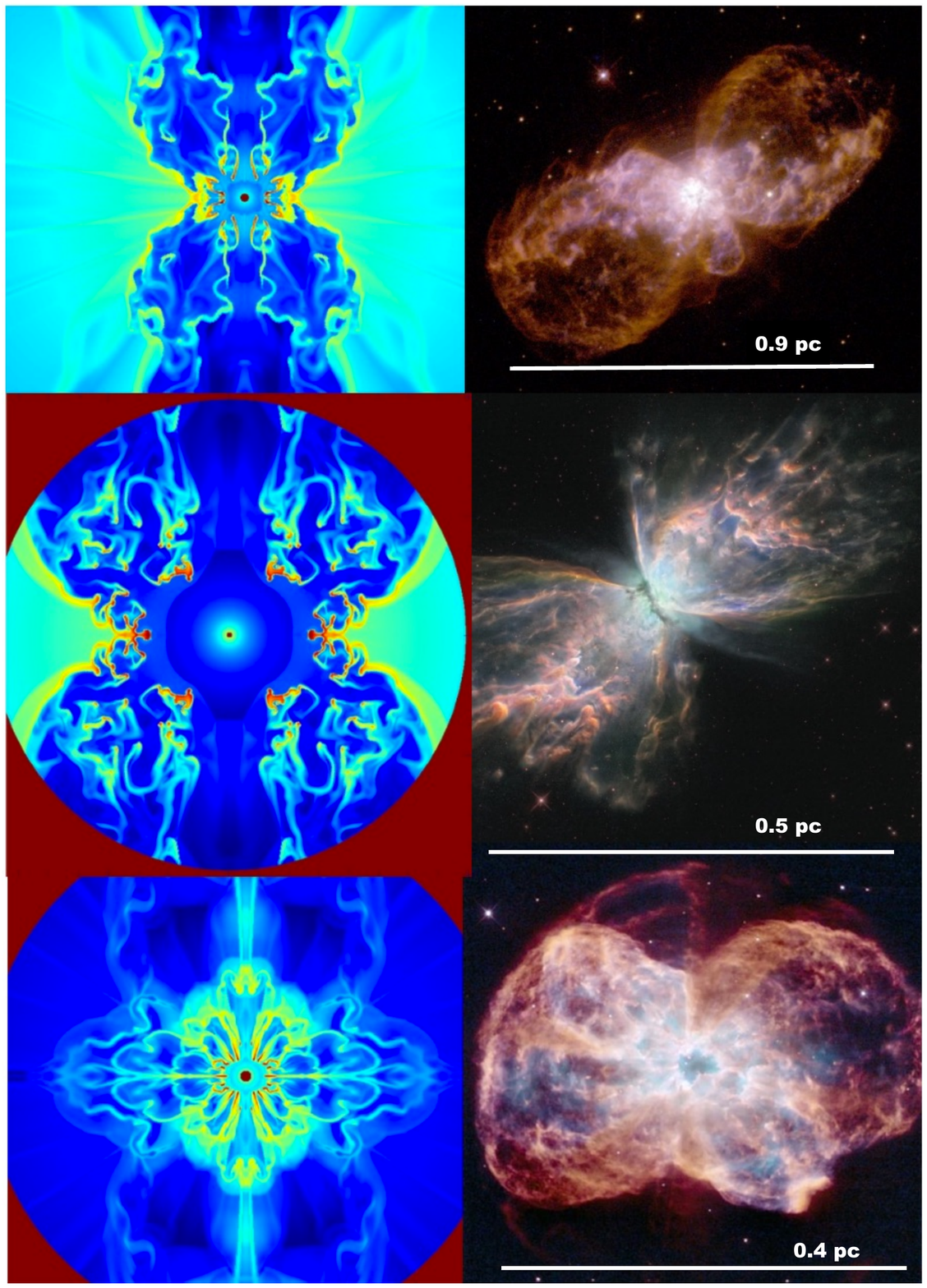}
    \caption{Examples of fragmentation and clumps. Left: density enlargement of models B6-2.5 (2,500 yr), C7-2.5 (2,500 yr) and C8-2.5 (3,000 yr). Right: Hubble 5, NGC 6302 and NGC 2440. Credit: NASA/ESA Hubble Space Telescope. }
    \label{F7}
\end{figure}

The primary
goal of this paper is to compute 
the type of ionized nebula resulting from 
a PPN that has undergone an 
active magnetic phase after a CEE event.
Note that this magnetic activity, produced by a circumbinary disk, is the result of a partial envelope ejection  during the CE phase. In the hypothetical case that the envelope ejection is totally
successful, the present results are inapplicable
since a circumbinary disk will never form.
In that case, the structure of the final PN should be 
similar to those models computed in Paper~I. 

 It should be pointed out that the degree to which the envelope is ejected 
during the CE phase has not been determined, although Ondratschek et al. (2022)  and Gonz\'alez-Bol\'{\i}var et al. (2022) have   
recently reported that total ejection is possible for a system with 
an AGB star where the recombination energy of hydrogen and helium can be tapped efficiently. 
In the case for a low efficiency of the ejection process, on the other hand, a massive circumbinary disk 
could form; however, a substantial amount of matter may return to within the binary orbit, which may facilitate 
merging of the system (Kuruwita el al. 2016). The properties of the binary system at the onset of the 
CE phase that we envision (yet to be determined) are such as to be bracketed by the systems leading to the above outcomes.

The models produce in general terms the typical morphologies observed in PNe, where 
most of the nebulae are elliptical (40.2\%), followed by the bipolar class (22\%) according to 
Gonz\'alez-Santamar\'{\i}a et al. (2021).  Our computations offer a new perspective for understanding  the formation of
elliptical nebulae  that has not been previously discussed in the context of CEE, 
 whereas bipolar nebulae have been attributed to binary interaction since the early work of Soker (1998).

The computations presented in this study show that bipolar nebulae can result from both low-mass and
intermediate-mass progenitors. This result implies that, according to the observed abundances, bipolar
PNe could be classified as type I or type II Peimbert (Peimbert 1978).
However, there is a difference in the resultant  morphology that could provide an indication of
the progenitor mass. The difference resides in the type of lobes produced. 
In particular,  central stars with larger masses have energetic winds at early phases, while low-mass progenitors
do not present important winds at the same  stage. This implies that larger masses
will produce more rounded lobes (Figure 2 and 3), while lower masses will produce more 
elongated and sharp-pointed shapes similar to the  PPN phase (Figure 4 and 5).
 The presence or absence (respectively) of diffuse  X-rays also follows.
 Since soft X-rays are  emitted by the hot shocked gas produced by the reverse
shock of the fast, line-driven stellar wind,  their detection indicates the existence of the fast wind (Kastner et al. 2000, Chu et al. 2001).

In addition, there is a difference in the resultant morphology that could provide a hint of
the progenitor mass.  The difference 
stems from the fragmentation of
the nebula, the origin of which resides in the temporal evolution of the wind.
Low-mass progenitors (1 \Mo) evolve very slowly  towards  the blue on the HR diagram, and
their respective winds start to be dynamically important only at very late times (10,000 yr). 
On the other hand, 
the ionizing radiation is important earlier in time. This dictates that the dynamics of the  nebula are
almost totally governed by the expansion of the \hii region, and the fragmentation is only based 
in the ionization-shock (I-S) front instability (Garc\'{\i}a-Segura \& Franco 1996). 
The I-S front instability occurs in  D-type ionization fronts (Kahn 1954)
and has a similar behavior as the thin-shell instability (Vishniac 1983).
 The thermal pressure in the thin-shell instability is provided by the hot shocked
gas from a wind-blown bubble or a supernova remnant, while the thermal pressure in the
I-S front instability is provided by the \hii region.

This type of fragmentation
produces cometary clumps aligned with the radiation field in the radial direction
(Models E-1.0 and F-1.0 in Figure 4). These cometary
clumps have been observed in a large number of PNe, for example in 
the Necklace nebula (Corradi et al. 2011)  and NGC 2392 (Guerrero et al. 2021).

In contrast, more massive progenitors (2.5 \Mo) produce hot shocked gas from the fast 
line-driven winds at early phases. The shocked wind, in combination with the radiation field, 
can fragment the nebula in a different form, as shown in Figure 6. Here,
the ionization front is very efficient in carving the neutral gas. 
The breaks made by the ionization front 
enable the hot gas to fill into these regions.
However, the hot shocked gas can expand in  the latitudinal direction very efficiently, 
and the resulting fragmentation  does not only occur in the radial direction. Thus, multiple clumps
are formed, with the difference that the tails of the clumps are not necessarily aligned in
the radial direction. Figure 7 shows a comparison of three models with the nebulae 
Hubble 5, NGC 6302, and NGC 2440. The three nebulae display multiple clumps within the
lobes  that are reproduced in the models B6-2.5, C7-2.5, and C8-2.5. 
Furthermore, model C8-2.5 produces an outflow in the equatorial region not aligned with the  major 
symmetry axis, produced by the  focusing of hot shocked gas  through the clumps. 
This may be similar  to that observed in NGC 2440 (L\'opez et al. 1998).  However, this  effect
could be an artifact {\bf of} the 2-dimensional computation, since the clumps here are indeed rings from 
the imposed symmetry, and
individual clumps could have a different behavior.

As a secondary goal of this paper, we checked whether or not the  magnetically collimated
outflows and jets computed in the PPN phase (Paper~III) were long-lived structures. 
Our computations show that those structures disappear as soon as the line-driven wind
is able to sweep up the collimated gas, leaving some remnants of  them in only a few cases.

Based on either observational work (Sahai \& Trauger 1998, Tafoya et al. 2020) or theoretical studies 
(Nordhaus \& Blackman 2006, Paper~III, Blackman 2022) 
jets have been suggested to form  
at the early stages of PPNe.
Based on the results of our simulations, we suggest that the observed jets in PNe
(Guerrero et al. 2020) must be remnants of those early phases and are not 
being actively collimated during the PN phases (Garc\'{\i}a-Segura 1997).
This implies that ``FLIERs'' and ``ansae''   (Balick et al. 1993, Gon\c calves et al. 2001) 
 could be remnants of those jets.

Figures 2, 3, 4 and 5 show that PNe  that form from PPNe   with large sizes (for example models B7, B8 and B9) contain dense and compact cores.  Note that the typical sizes of PPNe are of the order of 0.01~pc (Sahai \& Trauger 1998), while here the sizes are $\sim $0.2~pc . 
The presence of these small cores results from the lower expansion velocity of the CEE ejecta and the slow x-wind  (see Paper~III).
This effect produces dense gas close to  the center of the nebula (most likely unresolved in some cases). 
These compact cores can provide an explanation for the very bright center in
nebulae like the Red Spider Nebula (NGC 6537), 
 as in model B8-2.5 (Figure 2). 
We note that the bright cores could be confused with the total sizes of PNe that are very distant and 
have low surface brightness. These models could 
possibly be related to 
the PN morphological  class ``compact or stellar type''  and those unclassified  nebulae that are 
unresolved in surveys (Schwarz et al. 1992, Corradi \& Schwarz 1995, Manchado et al. 1996).

\section{Conclusions}

It has been found that models for PNe 
evolving from magnetically 
driven PPNe  that have  undergone a CEE event 
are able to explain the majority
of  PN shapes, i.e., the elliptical and  bipolar morphological classes. 
The point-symmetric class is not addressed in this paper, as such nebulae cannot
be studied in the two-dimensional approximation. Three-dimensional axis-free computations 
will be required to investigate this class.

 One and 2.5 \Mo (ZAMS)
progenitors can form bipolar PNe, but there are differences in 
the resultant nebulae. Specifically, larger masses
produce more rounded lobes, while lower masses produce more 
elongated and sharp-pointed shapes similar to the preceding  PPN phase.

There is also a large difference in the fragmentation of PNe as a consequence of the central stellar mass.
The fragmentation of low-mass PN progenitors produces cometary clumps aligned with the radiation field in 
the radial direction, mainly in the equatorial region producing ``necklace''-type nebulae.
On the other hand, larger masses 
give rise to the formation of multiple clumps in the latitudinal direction, 
which remain inside of the lobes, close to the center, 
are immersed  in and surrounded by hot shocked gas, and are not necessarily aligned in the radial direction.
 Although the large difference in the fragmentation is  mainly due to the
progenitor mass, it is also an explicit fingerprint of a post-CE PPN. 
In particular, the fragmentation is made possible by the gas distribution in the previous post-CE PPN as shaped by the jets.

The magnetically collimated outflows and jets that are formed in the  PPN phase disappear as soon as the 
line-driven wind is able to sweep up the collimated gas, leaving some remnants of their structure in the 
 PN phase.

There are still many unknowns in the formation of planetary nebulae descended from CEE. For example, the relationship between the mass ratios of the two stars and the ejection efficiency, the lifetime of the circumbinary disk,  and the role of the accretion disk  surrounding the secondary star 
 remain to be studied. 
The formation of point-symmetric nebulae still remains 
 challenging.
These open questions will be  part of our future work.

\section*{Acknowledgements}

 We thank our referee, Orsola De Marco, for a careful reading of the manuscript and for
her suggestions, which improved considerably the article.
We thank Michael L.\ Norman and the Laboratory for Computational
Astrophysics for the use of ZEUS-3D. The computations
were performed at the Instituto de Astronom\'{\i}a-UNAM at  Ensenada.
G.G.-S.\ is partially supported by CONACyT grant 178253.
Partial support for this work has been provided by NSF through grants
AST-0200876 and AST-0703950.

\section*{Data availability}

The data underlying this article will be shared on reasonable request to the corresponding author.



\bibliographystyle{mnras}


\bsp	
\label{lastpage}
\end{document}